# A bibliometric investigation into the literature of semantic reasoning in Internet of Things

Shayegan. M.J, University of Science and Culture

***Abstract*— Nowadays, semantic interoperability is a new keyword in the Internet of Things (IoT) for the exchange of information between sources. The constant need for interaction and cooperation has resulted in the creation of the Semantic Web with the help of tools and reasoners which manage personal information. Given the significance of the IoT and the increasing use of semantic techniques in this field, the present bibliometric investigation was conducted in the domain of semantic reasoning in the IoT. Bibliometrics involves analyzing bibliographic data of scientific sources, and it can be employed to arrive at an analysis of the status quo in a scientific field. In this study, through the analysis of 799 articles retrieved from the Web of Science database, distribution of topic categories, prolific and influential authors, language of articles, publishers of articles and their geographical distribution, the most debated/researched and the most frequently cited articles, and keyword trends were studied. The results of this study indicate that the number of articles published in the domain of semantic reasoning in the IoT has increased considerably in recent years. Of the articles analyzed, it was revealed that 10 countries produced 84% of the total documents, with China being in the lead. Moreover, as a result of keyword analysis, it can be maintained that the words fog computing, edge computing, Semantic Web, and wireless sensor network are among the most important keywords in this domain. As well, ontology has the highest average number of citations among the specialized keywords.**

*Impact Statement —* **The Internet of Things (IoT) is one of the most popular and trendy fields in computer science these days. Among the various fields of research that are being done on the IoT, the employing artificial intelligence techniques in the IoT is of interest to many researchers. Nowadays, it is necessary for a researcher to know the bibliometric information of his field. This bibliometric study evaluates the status of semantic reasoning in IoT research and provides researchers with useful bibliometric information.**

Mohammad Javad Shayegan is with the Department of computer engineering, University of Science and Culture, Tehran, Iran (e-mail: shayegan@usc.ac.ir )

## I. Introduction

Differences in the definitions of the Internet of Things (IoT) stem from the fact that it has various stakeholders including occupations and research studies. Everyone attempts to define and interpret the IoT according to their particular needs, concerns, and contexts. Noteworthy is also the fact that heterogeneous devices used in large IoT networks generate substantial amounts of real-time data which are transmitted to clouds through gateways for further processing [1]. Device heterogeneity can be due to capacity, features, sellers, and specific needs of the application. Although the Web is regarded as a suitable platform for integrating objects, the Semantic Web can enhance its capacity to understand object data and facilitate interoperability [2] . In addition to the heterogeneity of devices, technologies, protocols, data formats, semantics, hardware platforms, software packages, radio frequencies, and processing strategies are considered important aspects of the IoT physical environment. Therefore, data heterogeneity causes a lack of interoperability among applications and devices.

Semantics in the IoT refers to the ability to intelligently extract information employing different machines to provide necessary services. Semantic technologies can help realize such features as interoperability and reasoning in the IoT. However, because of the dynamic and heterogeneous nature of IoT data, limited sources, and real-time needs, using this technology is faced with a number of challenges [3] . Applying semantic technologies to the IoT enhances interaction and cooperation between IoT devices and facilitates data access, data integration, discovery of sources, semantic reasoning, and knowledge extraction.

The science of classifying and answering queries is at the core of Semantic Web reasoning services. Employing reasoners is one of the methods of reasoning knowledge in the Semantic Web. A semantic reasoner is a program which can deduce logical conclusions from a set of obvious facts or terms. A bibliometric study, this research centers on the existing articles in the Semantic Web.

Bibliometrics involves analyzing bibliographic data of scientific sources, and it can be used to reach an analysis of the status quo in a scientific field. These analyses include, among others, the identification of the most debated and/or researched topics in a certain field of science. Illustrating studies in a particular field, these analyses tend to be of immense interest to researchers, so their results are generally published as independent articles in international scientific journals. Drawing upon the Web of Science database, the present study is a bibliometric investigation into the domain of semantic reasoning in the IoT.

This bibliometric study, aimed at obtaining more accurate information in the above-mentioned domain, retrieved and analyzed the results of research done in scientific fields. The following are Sections 2, 3, and 4 which respectively present the method, results, and conclusions of this research.

## II. Method

The Web of Science (WoS) is one of the most commonly used scientific databases in bibliometric studies [4]. The only scientific database employed in this study was the WoS. To retrieve relevant articles in the field of semantics, as well as semantic reasoning, in the IoT, the researchers used the following as search terms:

"Internet of Things" and "semantic reasoning" or "reasoning"

As stated above, the study began by searching for articles on the WoS, a well-known index of scientific and bibliographic documents. In this search, the results were filtered by "article" document type. Studies show that searching for a topic does not always lead to the retrieval of all relevant publications. As such, abstracts or entire texts of articles should also be taken into account. The present study, therefore, considered searching for related topics, titles, keywords and abstracts.

Through the searching step, which took place on February 18, 2021, a total of 799 review articles in this domain were retrieved from the WoS citation database. In addition to using the information provided in these review articles, to analyze them, the researchers employed some bibliometric tools such as VOSviewer [5] and BibExcel [6] as well. More specifically, the BibExcel software package was used to arrive at the trends of the articles. Moreover, Microsoft Excel 2010 was used to extract and tabulate statistics in different fields and calculate correlations between variables.

It was mentioned above that searching the WoS database with the intended search terms resulted in the retrieval of 799 articles. These articles, in turn, received 9,512 citations, as shown in Figure 1. It can be concluded from the information displayed in Figure 1 that the citing trend of articles in this domain has been increasing exponentially from 2011 to 2020 such that the year 2020 witnessed the maximum number of citations, i.e., more than 2400 citations, and the same trend is expected to continue in 2021.

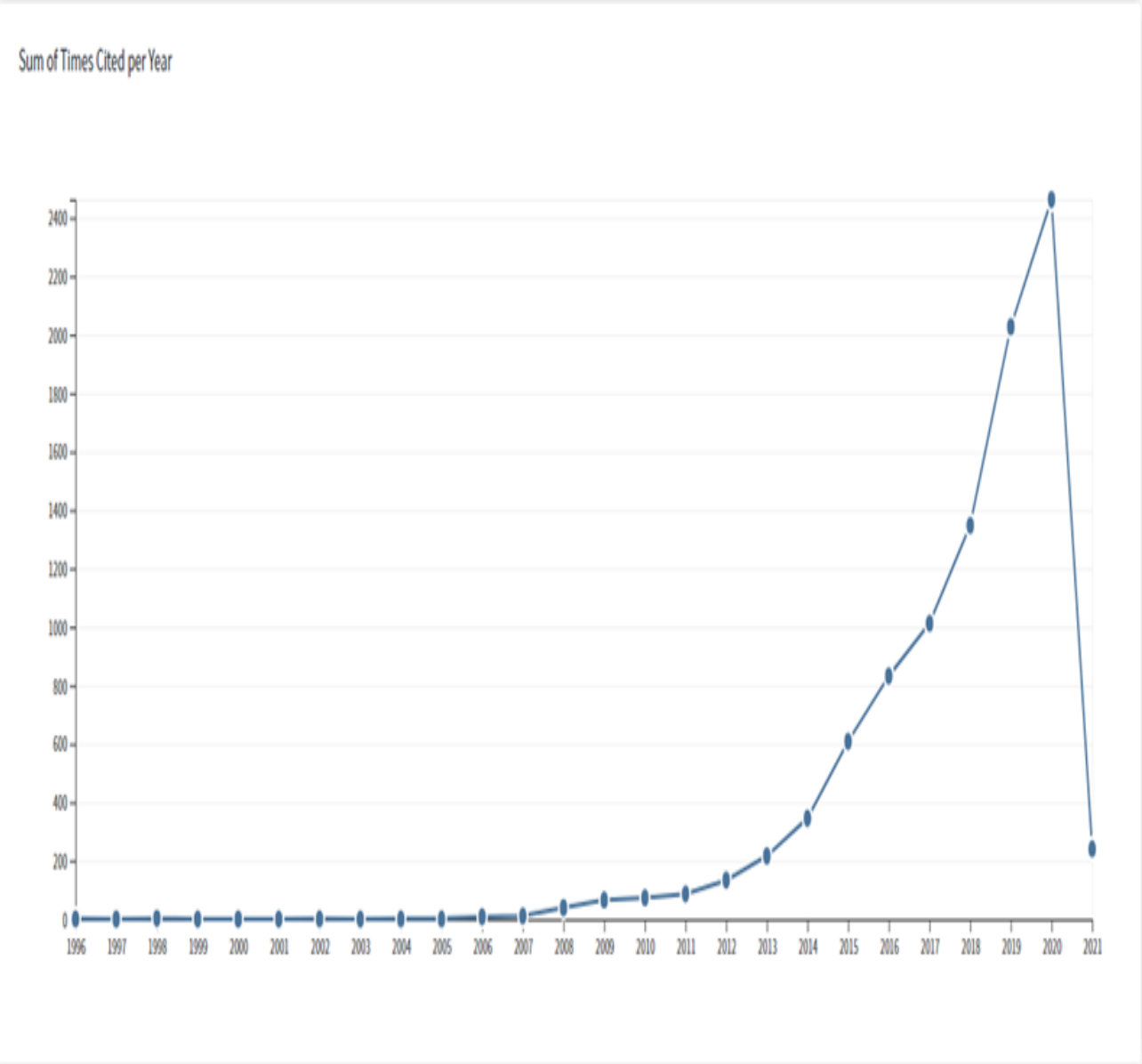

Figure 1: Number of citations in a year (based on WoS)

Every article available in the WoS database can be ascribed to at least one field category. Of all these articles, according to the WoS field categories, 34 percent are related to Computer Science, 26 percent relate to Electrical or Electronic Engineering, 16 percent pertain to Telecommunications, and the remaining 24 percent center around Business Economics, Chemistry, Instruments Instrumentation, etc. The corresponding illustration diagram is presented in Figure 2.

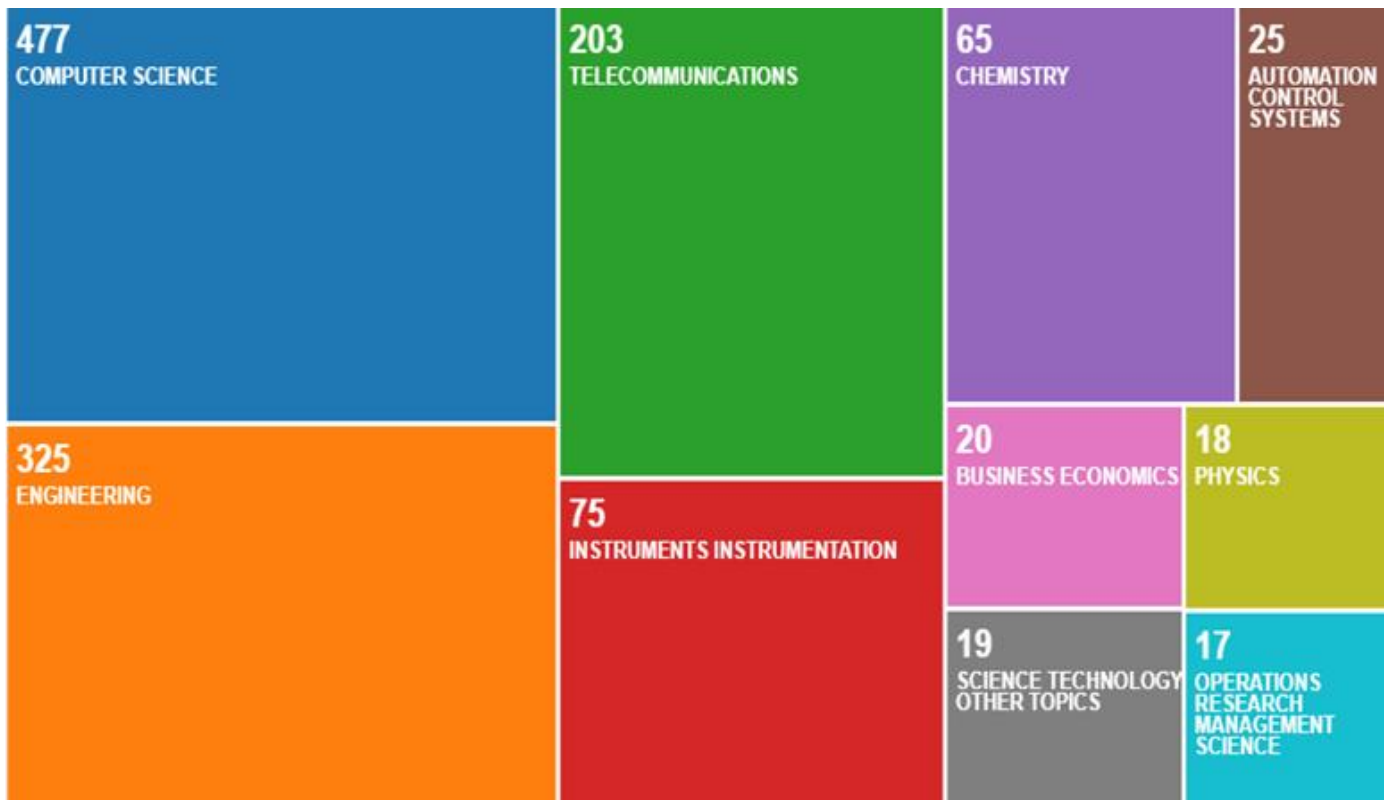

Figure 2: Illustration of 10 fields based on retrieved articles from WoS

The frequency of articles in the domain of semantic reasoning in the IoT is displayed in Figure 3. Likewise, this information was obtained from the WoS citation database.

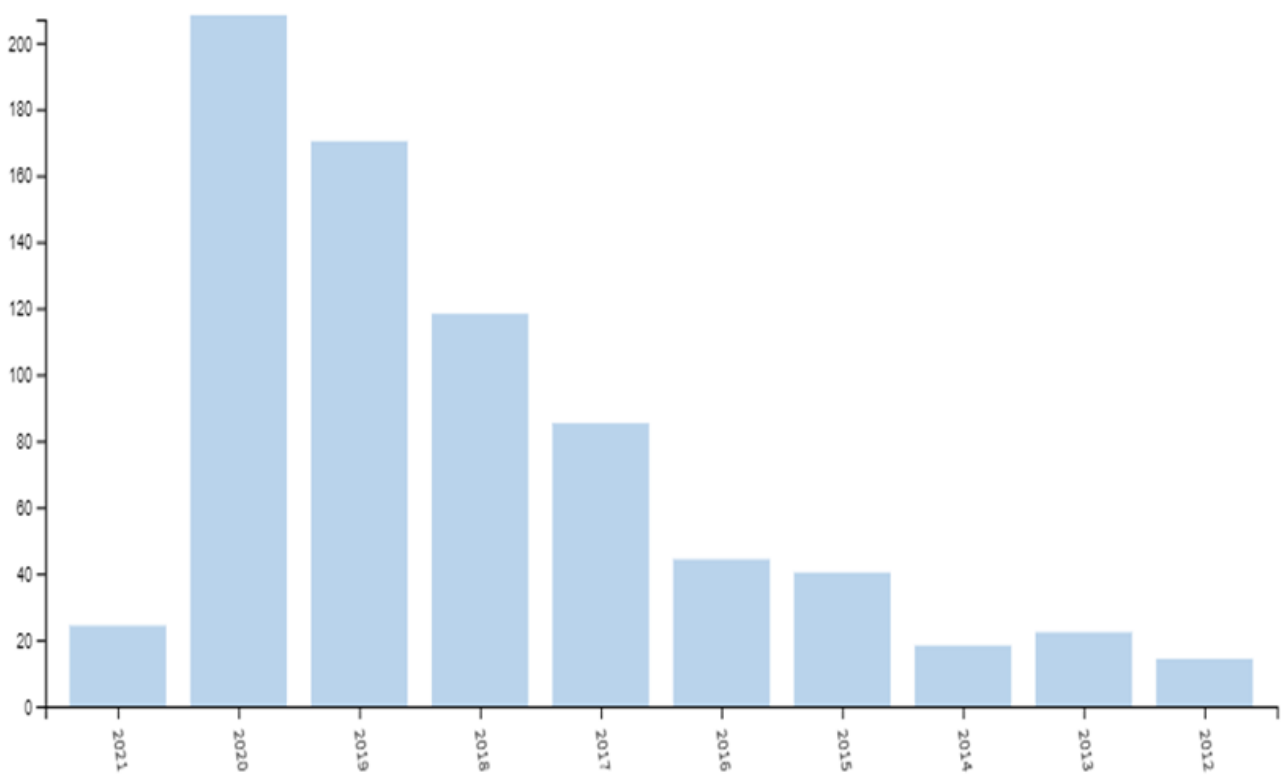

Figure 3: Frequency of articles published in the domain of IoT semantic reasoning between 2012 and 2021 (based on WoS)

According to the diagram in Figure 3, it can be perceived that the frequency of articles published in 2020 has been the highest number since 2012. In general, the growing trend of publishing such articles reveals researchers' tendencies towards this domain.

In the following section, statistical results representing prolific and influential authors in this domain, frequently cited articles, as well as keywords and other information about these articles, are provided and analyzed.

## III. Results

### A. Language of articles

First, statistics on the language of articles are provided. Subsequently, based on these statistics obtained through the software introduced in the previous section, the information retrieved from the WoS citation database is analyzed in more detail. According to the statistics, English is the dominant language of these articles, written in the domain of IoT semantic reasoning (see Table 1).

Table 1: Language frequency of articles in the domain of semantic reasoning in IoT

| Language | Frequency | Percent |
|---|---|---|
| **English** | 789 | 98.748 % |
| **Spanish** | 3 | 0/38% |
| **Portuguese** | 2 | 0/25% |
| **Russian** | 2 | 0/25% |
| **French** | 1 | 0/13% |
| **Korean** | 1 | 0/13% |

### B. *Prolific and influential authors*

Through the analysis of the statistics with the *Bibexcel* statistical analysis software, the 10 most prolific and influential authors in the domain of semantic reasoning in the IoT were identified (see Table 2). These authors' ranking was determined based on the H-index which is a bibliometric index essentially devised for evaluating authors. This evaluation is conducted through combining quantitative and impact-related data. Therefore, this study only used the H-index to analyze the citations to articles in this domain.

Table 2: 10 most prolific and influential authors in the domain of semantic reasoning in IoT

| ID | Author | All citations | All articles | h-index |
|---|---|---|---|---|
| 1 | Blanco-F Y | 334 | 17 | 10 |
| 2 | Gil-Solla A | 298 | 15 | 10 |
| 3 | Lopez-N M | 334 | 17 | 10 |
| 4 | Pazos-Arias J | 309 | 14 | 10 |
| 5 | Ramos-C M | 274 | 14 | 9 |
| 6 | Garcia-Duque J | 178 | 9 | 8 |
| 7 | Skarmeta AF | 326 | 11 | 8 |
| 8 | Fernandez- V | 150 | 8 | 8 |
| 9 | Diaz-R R | 126 | 6 | 6 |
| 10 | Jara AJ | 304 | 6 | 5 |

In addition to identifying these 10 most prolific and influential authors, the researchers arrived at the relation graph of the authors of the 799 retrieved articles. This graph was drawn using the *VOSviewer* software and is shown in Figure 4 below.

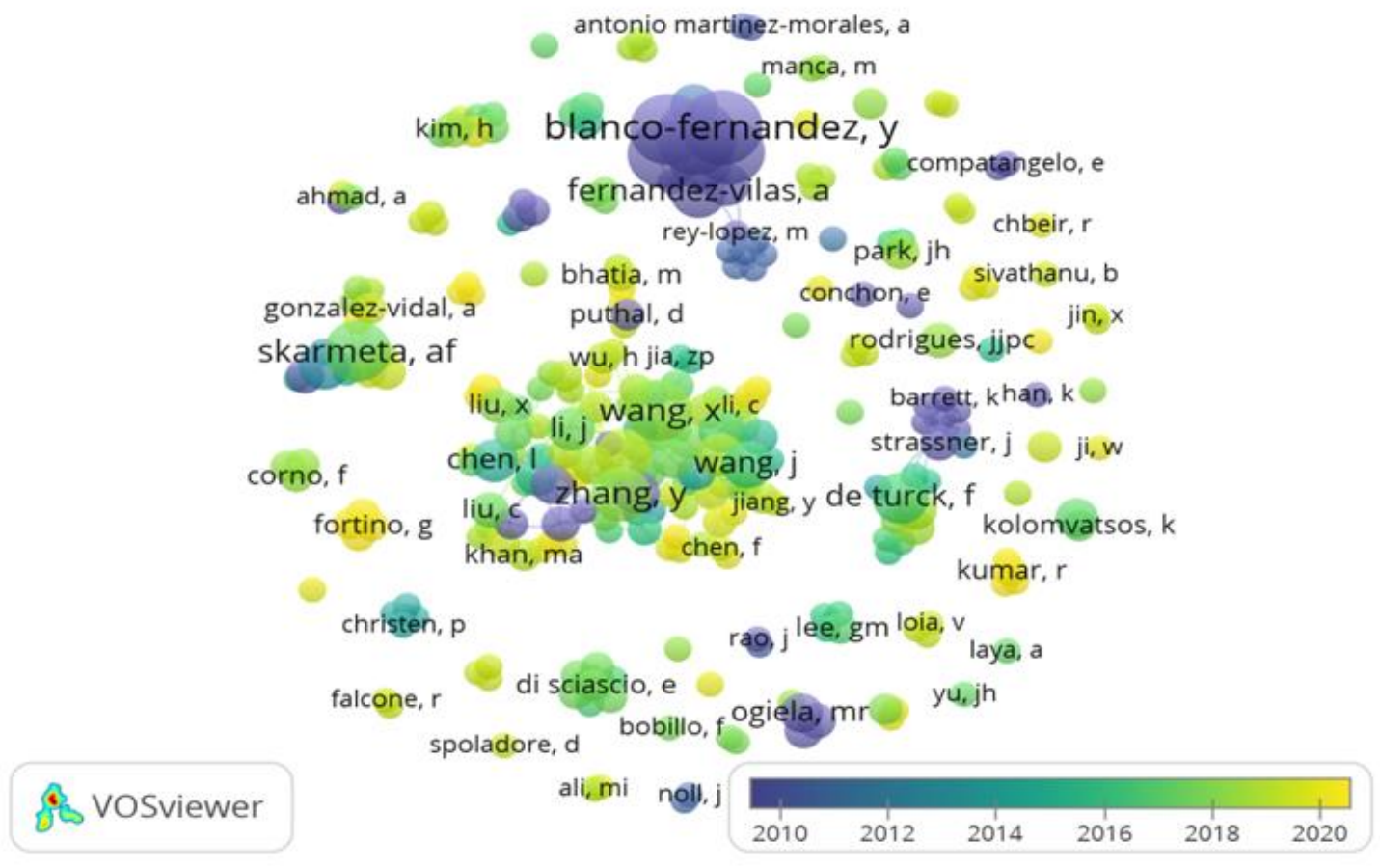

Figure 4: Authors' relations based on retrieved articles

As can be seen in Figure 4, the nodes in yellow show the authors who cooperated in 2020. Using this and the other colors, one can analyze in which year(s) the authors have been more prolific. Furthermore, the nodes which are bigger in size represent the authors who have been more prolific and cooperated more with other authors. Take Wang, Blanco-Fernandez, and Skarmeta as cases in point.

### C. *Most prolific universities*

Table 3 presents the most active universities in terms of publishing articles in this domain. With respect to Table 3, it can be concluded that the most prolific university in this domain has been the University of Vigo in Spain.

TABLE 3: 5 MOST PROLIFIC UNIVERSITIES IN THE DOMAIN OF SEMANTIC REASONING IN IOT

| ID | University | Country | Document |
|---|---|---|---|
| 1 | University of Vigo | Spain | 4 |
| 2 | Campus University | Spain | 3 |
| 3 | Yonsei University | South Korea | 2 |
| 4 | University of Glasgow | Scotland | 2 |
| 5 | Beijing Jiaotong University | China | 2 |

Nevertheless, although this university ranks first because of its considerable number of articles, the analysis of the 799 retrieved articles from the WoS database reveals that Chinese universities and organizations, in total, have been most prolific in the domain of IoT semantic reasoning (see Figure 5).

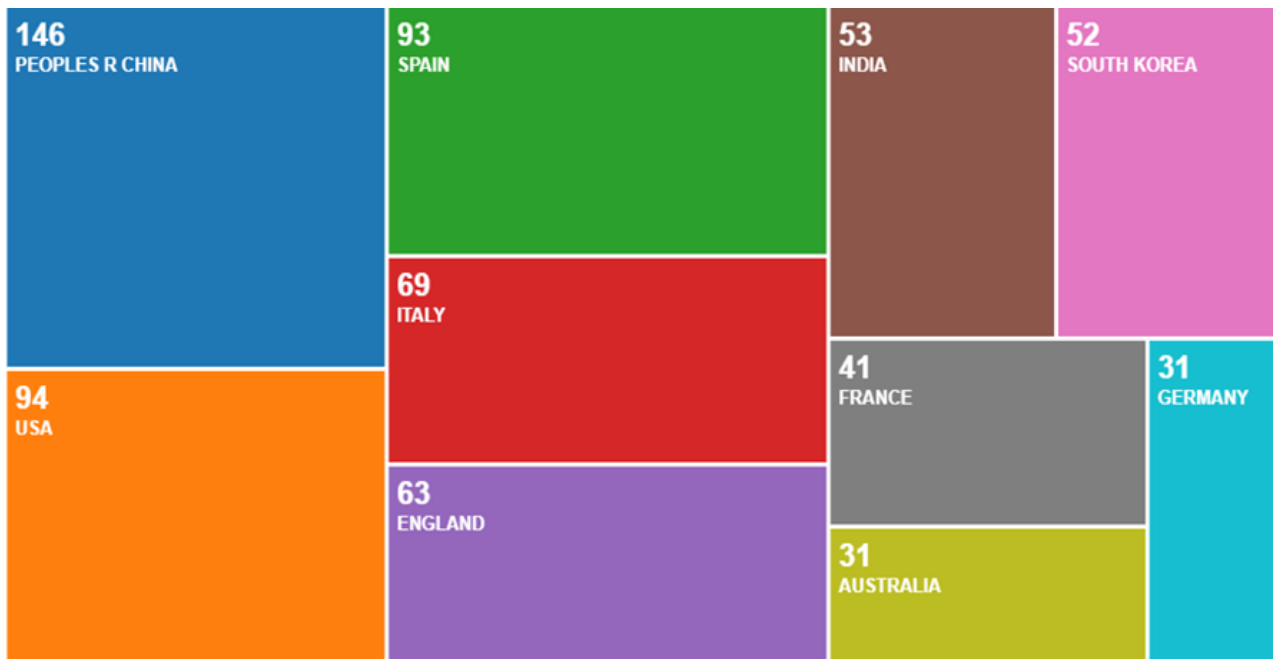

Figure 5: 10 most prolific countries in the domain of IoT semantic reasoning

In regard to the statistics in Figure 5, one may compute that approximately 18.27 percent of these articles belong to China, 11.76 percent are attributed to the United States of America, and 11.63 percent have been written in Spain. In turn, it can be calculated that nearly 84.23 percent of the articles belong to the 10 most prolific countries in this domain.

## D. *Most frequently cited articles*

The 10 most frequently cited articles in the domain of IoT semantic reasoning are presented in Table 4, implying that studying these documents by all researchers intending to do investigations in this domain is probably essential. The number of citations to these articles exceeds 100. Among these, a survey article by C Perera (2013) referred to 1136 times has been the most frequently cited one.

## E. *Most important keywords*

Authors tend to provide keywords as summaries of articles [17] . Therefore, first, the most frequent keywords used in the domain of semantic reasoning in the IoT were analyzed. Following this, trends representing the keywords were examined. The keywords were studied in three periods. Figure 6 presents the most important keywords obtained by means of the *VOSviewer*.

TABLE 4: 10 MOST FREQUENTLY CITED ARTICLES IN THE DOMAIN OF IOT SEMANTIC REASONING

| Id | Document Title | Citations |
|---|---|---|
| 1 | Context Aware Computing for The Internet of Things: A Survey [7] | 1136 |
| 2 | The ChEBI reference database and ontology for biologically relevant chemistry: enhancements for 2013 [8] | 337 |
| 3 | Towards the Implementation of IoT for Environmental Condition Monitoring in Homes [9] | 335 |
| 4 | A Knowledge-Driven Approach to Activity Recognition in Smart Homes [10] | 275 |
| 5 | Closed-loop PLM for intelligent products in the era of the Internet of things [11] | 233 |
| 6 | A Survey on Distributed Topology Control Techniques for Extending the Lifetime of Battery Powered Wireless Sensor Networks [12] | 195 |
| 7 | An internet of things-based personal device for diabetes therapy management in ambient assisted living (AAL) [13] | 143 |
| 8 | Interconnection Framework for mHealth and Remote Monitoring Based on the Internet of Things [14] | 115 |
| 9 | BIM and ontology-based approach for building cost estimation [15] | 111 |
| 10 | Sensor Search Techniques for Sensing as a Service Architecture for the Internet of Things [16] | 105 |

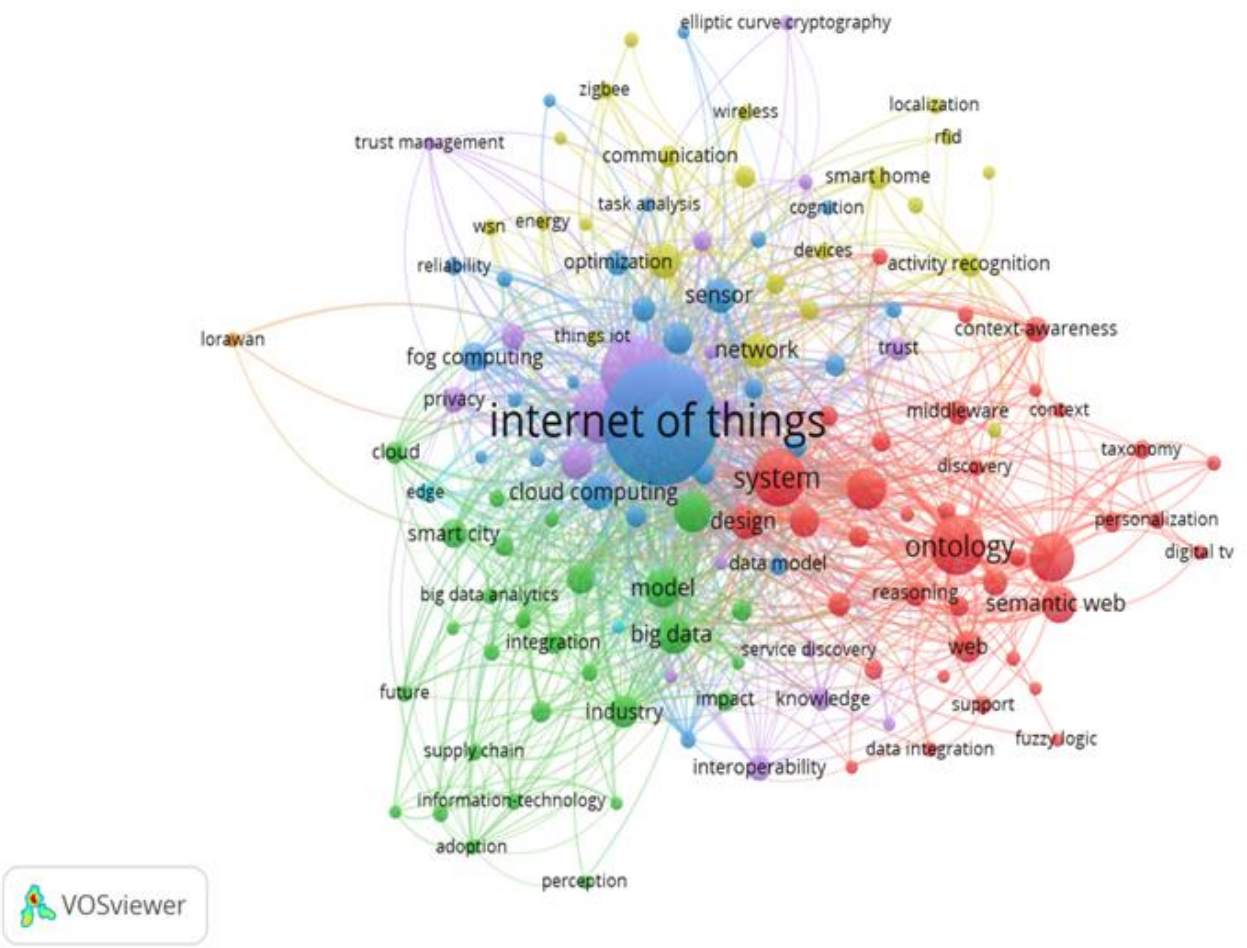

Figure 6: Schematic presentation of keywords in IoT semantic reasoning in VOSviewer

TABLE 5: TWENTY MOST FREQUENT KEYWORDS BETWEEN 2018 AND 2021

| Label | Weight <Links> | Weight <Total link strength> | Weight <Occurrences> | Score <Avg. pub. year> | Score <Avg. citations> |
|---|---|---|---|---|---|
| internet of things | 129 | 1113 | 375 | 2018 | 12 |
| Internet | 109 | 538 | 133 | 2018 | 11 |
| System | 103 | 315 | 81 | 2017 | 22 |
| Things | 92 | 402 | 85 | 2018 | 15 |
| Model | 82 | 185 | 39 | 2018 | 5 |
| Management | 80 | 199 | 42 | 2018 | 11 |
| Ontology | 76 | 270 | 87 | 2016 | 24 |
| Framework | 75 | 184 | 46 | 2017 | 33 |
| big data | 68 | 156 | 36 | 2018 | 9 |
| Design | 68 | 130 | 29 | 2017 | 15 |
| Security | 67 | 180 | 46 | 2018 | 8 |
| Challenges | 66 | 143 | 28 | 2018 | 16 |
| Sensor | 58 | 109 | 32 | 2018 | 15 |
| wireless sensor network | 56 | 133 | 32 | 2017 | 20 |
| Industry | 54 | 121 | 25 | 2019 | 8 |
| cloud computing | 53 | 125 | 33 | 2019 | 8 |
| smart city | 53 | 94 | 22 | 2018 | 9 |
| Information | 53 | 75 | 14 | 2016 | 100 |
| Network | 52 | 110 | 29 | 2018 | 14 |

As can be seen in Tables 5, *Internet of things* is the most frequent keyword (weight 128) in the related works. Apart from some general keywords, the frequency of *ontology* is significant (weight 76). This matter shows that a major part of research methods on semantic reasoning were performed by *ontology*. The frequency of *big data* (weight 68) can be interesting among the researchers. Another interesting point in Table 5 is related to the *score of average publication year* column, where “industry” and “cloud computing” have the highest rate. For *average number of citations* column, among the specialized keywords, *ontology* has the highest score.

The corresponding keyword cloud is displayed in Figure 7. Keyword clouds are often used to visually summarize textual documents [18]. The size of each keyword in the cloud denotes its relative frequency or usage.

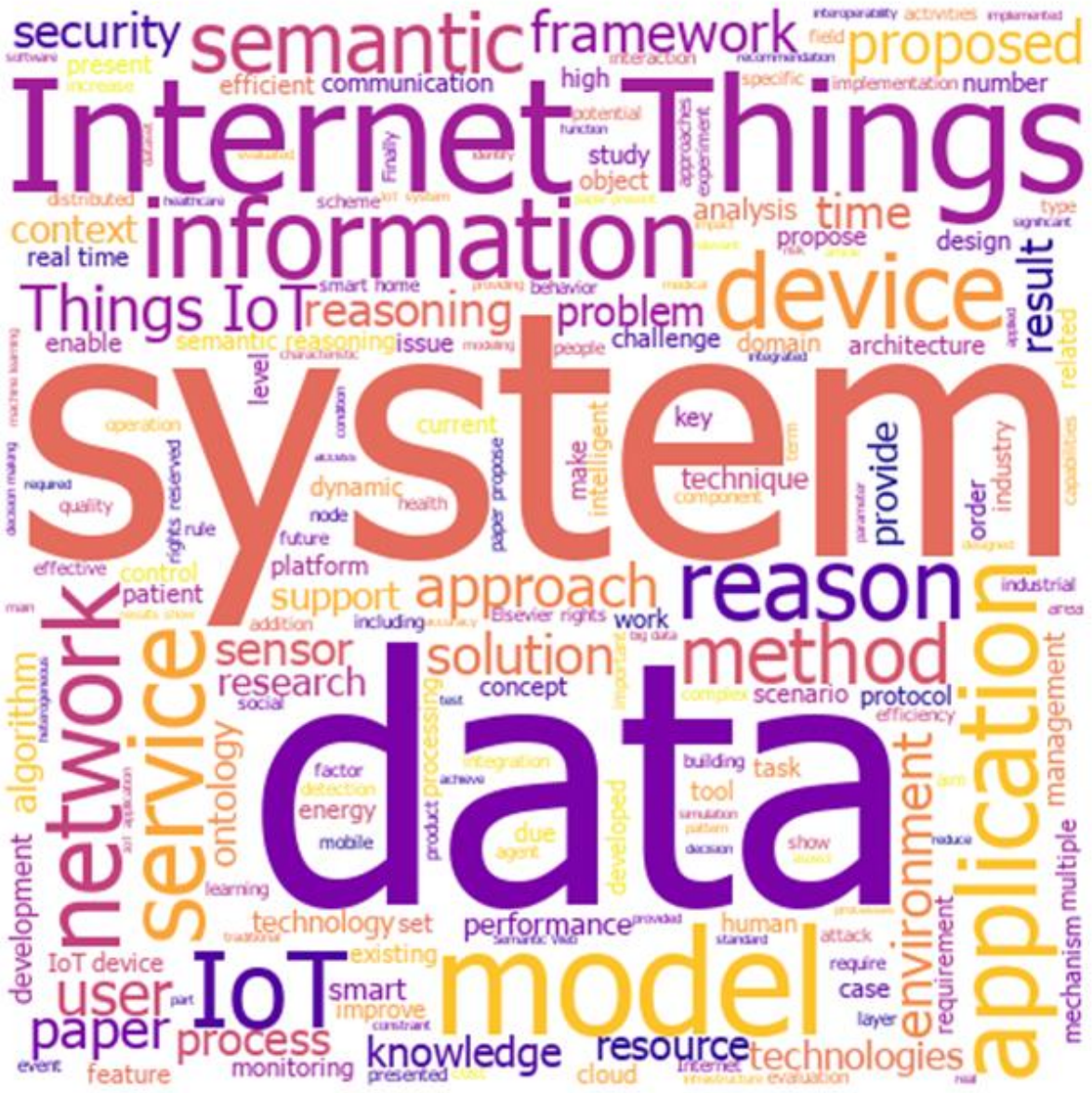

Figure 7: Keyword cloud used in IoT semantic reasoning

With respect to the keyword cloud derived based on the abstracts of the 799 retrieved articles, it can be concluded that the most frequent keywords in these abstracts are *data*, *system*, *Internet of Things*, and *reason* which, given their high frequencies, are displayed in larger sizes than the other less commonly occurring terms.

## IV. CONCLUSIONS

The current research presented a bibliometric investigation in the domain of semantic reasoning in the Internet of Things. The analysis conducted in this study shows that the year 2020 observed the publishing of the highest number of documents (that is, 191 articles) in the domain of semantic reasoning in the IoT. Additionally, of the documents analyzed, 10 countries produced approximately 84% of the total, with Chinese universities and organizations being the most prolific ones in the IoT semantic reasoning domain. The most articles published in this domain, however, belong to the University of Vigo in Spain. Furthermore, the analysis of keywords revealed that in addition to the term *Internet of Things*, the most widely used keywords are *fog computing*, *edge computing*, *Semantic Web*, and *wireless sensor network*, considered to be the most significant terms. More investigation showed a major part of research methods on semantic reasoning were performed by *ontology*. As well, *ontology* has highest average number of citations among the specialized keywords. This study was conducted on WoS citation database. However, according to related researches on WoS and Scopus, the results of bibliometric investigations into natural sciences and engineering conducted based on WoS and Scopus have noticeable overlap.

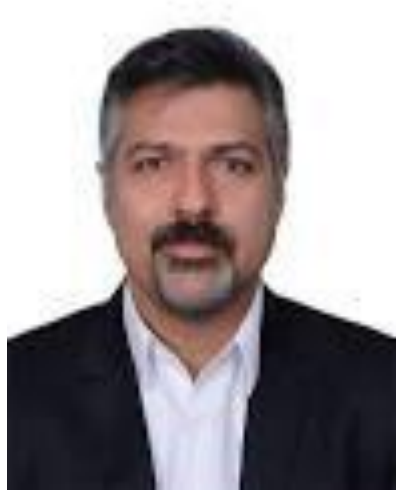

**Mohammad Javad Shayegan** is an Associate Professor at the Department of Computer Engineering at the University of Science and Culture, Tehran, Iran. He received his Ph.D. degrees in Information Technology and Multimedia Systems from the University Putra Malaysia in 2008. He is the founder and program chair for the IEEE International Conference on Web Research (ICWR) and Editor-in-Chief for the International Journal of Web Research. His research interest is Web and data science, and distributed systems.